**Personal Recollection**

# Chronicle of the discovery of the back-bending phenomenon in atomic nuclei: a personal recollection 50 years on

Hans Ryde[a]

Department of Physics, University of Lund, Lund, Sweden

**Abstract** A chronicle describing the historical context and the development of ideas and experiments leading to the discovery of the back-bending phenomenon in rapidly rotating atomic nuclei some 50 years ago is presented. The moment of inertia of some atomic nuclei increases anomalously at a certain rotational frequency, revealing important clues to our understanding of nuclear structure. I highlight the decisive interactions and contacts between experimentalists and theorists, which created the right environment, allowing for the revelation of an undetected phenomenon in Nature. Finally, I reflect on the key points allowing for the discovery and particularly point to the importance of systematic surveys, which in this case investigated the energy levels in heavy nuclei of a large sample of elements, as well as to the accuracy of the measurements of the ground state levels made at the time.

## 1 Introduction

Advances in science may happen in many ways, sometimes they are planned but often they occur unexpectedly. The investigation, by which the back-bending phenomenon was serendipitously discovered, was a planned endeavour. The purpose was to make a survey for a large range of atomic nuclei of how their energy levels and spins change with the angular momentum introduced into the systems. During this investigation, we unexpectedly discovered that the moment of inertia of some elements increased anomalously at a certain rotational frequency, a property that is now denoted as back-bending.

I was inspired to start this investigation in a similar way as Bengt Edlén[1]—my esteemed professor at the physics department at Lund University—once followed in his studies of atomic spectra and energy levels of atoms. Edlén started by collecting all available relevant data on atomic energy levels in a summer-empty Uppsala physics laboratory in about 1930. Based on these data, and his own new measurements, he could then conclude that the energy levels of atoms followed the shell structure which changed with atomic number or with the charge of the nucleus. This result was later used to solve the Coronium mystery. Some unknown spectral lines observed in the solar spectrum were first suggested to be associated with a new element, dubbed Coronium. Edlén now understood that they were in fact emitted from already known, highly ionised elements.

In a similar way, I started a systematic investigation of the energy levels in heavy nuclei at the Nobel Institute for Physics in Stockholm after my return from my post-doctoral years at the California Institute of Technology in 1965. During a visit to the nuclear chemists Dick Diamond and Frank Stephens at the Berkeley National Laboratory in California, I had learned how to study heavy nuclei by exciting them by using accelerated $\alpha$-particles. On my return to Stockholm, I realised the possibility of this method to perform a similar analysis to Edlén's, but this time with atomic nuclei instead.

Consequently, all the information on atomic nuclei available at the laboratory of the Nobel Institute for Physics in Stockholm was gathered, and their energy levels were systematically recorded, using the beam of 43 MeV $\alpha$-particles extracted from the newly rebuilt 225-cm Stockholm cyclotron. We studied a range of nuclei with atomic numbers between about 100 and 200 in ($\alpha$, xn) reactions. In particular, we analysed the dependence of the nuclear energy levels on the rotational excitation energy.

---

[1] Bengt Edlén (1906–1993) professor of physics at Lund University.

[a] e-mail: hans.ryde@nuclear.lu.se (corresponding author)

## 2 Unexpected deviations and discussions with theorists

In the course of the investigation, we noticed some deviations in the spectra of the $\gamma$-ray lines in the rotational bands. For instance, in the paper [6], we reported on a measurement of the $^{160}$Dy$(\alpha, 3n)^{161}$Er reaction. The rotational band we observed for $^{161}$Er deviated somewhat from what was expected with increasing spin. This was indeed one of the earliest hints of a change of rotational behaviour of nuclei at high spins. The full recognition of how important the observations were, still lay a few years ahead. To try to find an explanation for this, I turned to the leading nuclear structure theorists Aage Bohr and Ben Mottelson in Copenhagen, with whom I had had previous contacts regarding theoretical interpretations of our experimental data. Bohr and Mottelson had made recent progress in the theory of nuclear structure and were therefore the obvious experts with whom to discuss these new results (e.g. [4,13]).

At a nuclear physics meeting arranged in Jönköping in southern Sweden, I met Aage Bohr and discussed with him, among other things, our new data on the rotational energies in the bands in $^{155}$Gd as well as bands in the doubly-even Hf nuclei. Bohr was particularly interested in the ground-state rotational band in $^{162}$Hf. After the Jönköping meeting, in a letter dated 5 February 1969, Aage Bohr mentioned that he and Ben Mottelson were very interested, in particular, in the results of our measurements of the ground-state rotational band, which they noted, represented a significant step forward. He also suggested that we should show the moment of inertia as a function of the square of the frequency, that is $\omega^2$. This is what is typically done in theoretical calculations within the "cranking model". This way of presentation gives a better picture of what happens within the nucleus and indeed is now the standard way of plotting the relation.

In the course of further investigations, we also found particularly interesting results on the nucleus of $^{160}$Dy. We again sent our manuscript to Bohr and Mottelson. They acknowledged the importance of our results in a letter dated 16 October 1970. In this letter, they used their suggestion from early 1969 and plotted our data of the inertial momentum for $^{160}$Dy as a function of $\omega^2$. The hand-drawn plot in their letter is shown in Fig. 1, which also shows a curve having a clear "up-turn". This is the first representation of the behaviour that would soon be denoted back-bending. A week after receiving the letter, we submitted our paper on $^{160}$Dy, including the figure that they had suggested [9]. Arne Johnson was one of the many graduate students in my group at the time, and was the one assigned responsibility for the data analysis of the $^{160}$Dy nucleus [2]. He later defended a thesis on this subject in 1973. His contribution to this work is evidenced by him being invited to write the review [11]. Josef Sztarkier was the technical expert who set up and developed the coincidence unit, which was the decisive feature of the experiments, allowing us to accurately identify the reactions. Following papers include [3,10] and [17]. Here, more nuclei were presented, including $^{158}$Dy, $^{160}$Dy and $^{162}$Er, $^{168}$Yb, with the full S-curve of $^{162}$Er which is clearly visible, showing the archetypal back-bending behaviour.

## 3 Early interpretations and developments

In connection with these developments, Stephens and Simon submitted a paper in November 1971 that presented a detailed theoretical work on the Coriolis effects at high angular momentum in nuclear systems [19]. One aim of this paper was to explain the experimental observations that we had made, first in [6], and later in [9]. Among other things, they discuss the formation of the S-shape curves of the back-bending phenomenon that were being observed experimentally.

Moreover, Bohr and Mottelson suggested that the S-curve could be due to an abrupt change in the structure of the nucleus, maybe a change in the symmetry. At low angular momenta, the nucleons keep together in pairs. At higher angular momenta, the rotational energy exceeds the energy needed to keep these pairs together and they split. This process costs energy, leading to the detected change in moment-of-inertia of the spinning nucleus. Another explanation was suggested, namely a phase transition of the nuclear matter from a superfluid state to a state of independent particle motion. The pairing collapse will cause an increase in the moment of inertia and only a smaller one in spin. Indeed, the concept of back-bending was introduced in 1960 by Mottelson and Valatin as a similarity between the effect of rotation on the nuclear pairs and the effect of the magnetic field on electron pairs in a superconductor. Thus, there is a force that acts oppositely on the pair and weakens the pairing correlation and causes a sudden collapse.

The response to the discovery was unexpectedly large and I was invited by Professor Sugihara to the 1971 Gordon Research Conference on Nuclear Chemistry which took place 21–25 June 1971 in New Hampshire, USA. The response was overwhelming and I remember the queue of enthusiastic colleagues who wanted to see the plots of the results.

An important person in this development was Zdziław Szymański (cf. [15]), a Polish theoretical physicist periodically working at the Niels Bohr Institute in Copenhagen at the time. When I was a post doc. at Caltech in 1963–65, I shared an office with Zdzisław at the Norman Bridge Laboratory as members of Felix Boehm's group. It was here

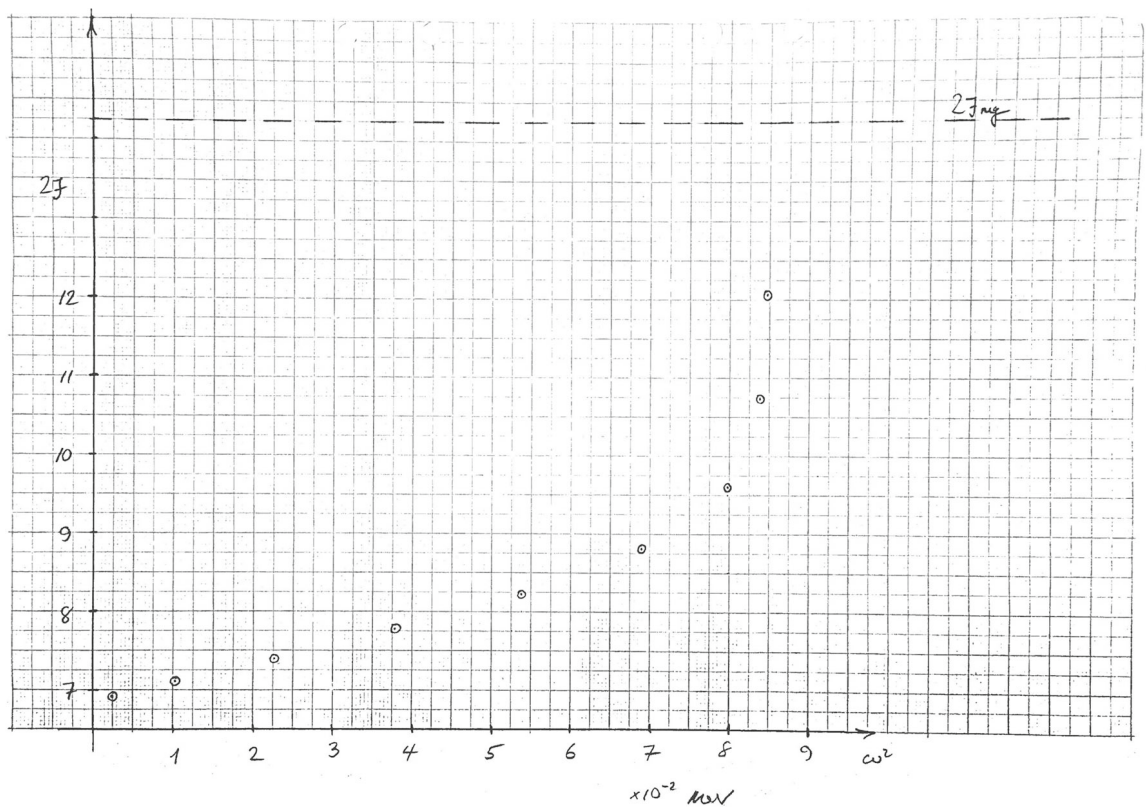

**Fig. 1** Bohr and Mottleson's original plot of our data for $^{160}$Dy, in which the angular momentum is plotted versus $\omega^2$ for the first time. This figure later appeared in [9]

that I learned not only most of what I know about nuclear structure theory from the discussions with Zdziław, but also how to understand the lectures of Richard Feynman and Murray Gell-Mann, which were being given at that time. It was most fortunate that both Zdzisław Szymański and Ray Sorensen of the University of Pittsburgh (USA) happened to be spending several months at the Research Institute for Physics in Stockholm in 1970–1971. The subsequent observations and interpretations of intricate spectra of gamma rays with many rotational bands were made possible through the important collaboration with them. The visit of these two eminent physicists created the stimulating environment for the further investigation of the back-bending phenomenon. Szymańsk's contribution played a decisive role in giving the theoretical explanation acceptance.

When we pondered over our graphs of the moment-of-inertia as a function of the square of the rotational frequency, which showed a characteristic S-type shape, I remember that Zdzisław Szymański likened their shape to the form of the bow of a viking ship. Other early working names for the phenomenon described by these graphs were the S-curve, the Superband, or the Stockholm band. Eventually, it came to be known as the back-bending phenomenon.

In 1972, a Symposium on High-Spin Nuclear States and Related Phenomena was arranged at the Research Institute for Physics in Stockholm [18]. The discovery of back-bending had opened up a barrage of activity and research in groups around the world, with experiments all trying to break up pairs of nucleons in even–even nuclei by exciting them to high rotational energies. The symposium clearly showed that nuclear structure had become the epicentre of interest in nuclear physics.

## 4 Gamma ray detector techniques and accelerators

An important step ahead for this research was the replacement of the classical crystal material NaI(Tl) as a scintillation detector. When I worked at the California Institute of Technology in Pasadena 1963–65, I started using a novel crystal material, namely Li-drifted germanium (Ge). Such crystals provided a much better energy resolution and counting efficiency because of the higher density and smaller band gap, leading to a much better energy resolution. The naked, Li-drifted Ge(Li) crystals of a few cm$^3$ were made by the small company in Santa Monica, CA, and I drove over to them, some 30 miles, to get a few test crystals. I remember that I had to keep them cold, dry, and with a reflecting aluminium cover, during my drive back to the laboratory at Caltech. The first experiments we performed using this new crystal material was in a study of the gamma ray spectrum from the decay of the radioactive $^{181}$Ta nucleus [1]. With the higher-resolution spectrum, we found a doublet ground state with a 6 keV energy difference. The subsequent discovery of back-bending was, in part, made possible due to this new detector material.

Further important experimental progress was the development and the use of more advanced accelerators to reach higher energies and utilizing the better energy resolution. The old 225-cm cyclotron that was used for the back-bending discovery experiment produced a maximum energy of about 10 MeV per nucleon, using $\alpha$-particles. When we changed to using $^{12}$C -ions, we could instead reach the maximal energy of 120 MeV. We later used the Danish tandem accelerator at the Niels Bohr Institute at Risø, a few miles west of Copenhagen, where we could get 10 MeV per nucleon for ion sources of different $Z$ elements.

Apart from the energy of the bombarding ions, the experimental techniques for the investigations were also developed. For nuclear reactions, the angular distribution of the gamma rays is important for the classification of the radiation. Initially, we had set up systems in the ($\alpha$, xn) experiments in Stockholm to measure the angular distribution of the gamma radiation. The germanium detectors were moved at an angle relative to the incoming beam. The discovery of back-bending illustrated clearly the importance of an experimental setup that is able to measure the angular distribution of gamma rays. This encouraged the development of multi-detector systems. After I shifted to the university of Roskilde in Denmark 1972, we started to work with the Risø tandem accelerator and various ion sources. At Risø, Ge(Li) detectors were developed, with the construction of multi-detector systems in order to be able to measure the angular dependence of the reaction products.

Bent Herskind and collaborators at the Niels Bohr Institute initiated a flexible multi-detector system for studying nuclear reactions with various heavy ion beams. This was the introduction of the NORDBALL project, which was a collaboration among mainly Nordic physicists [7]. At the University of Lund, to which I moved in 1976, we started to make a mechanical structure to get the Ge(Li) detectors at various angles and distances to the ion beam. The structure, which was finally selected, had the shape of a football, mathematically known as a truncated icosahedron, in which the Ge detectors could be mounted. As part of the design study, my twin sons Nils and Felix built a full-scale model of the truncated icosahedron in cardboard, glue, and tape, at home in the kitchen. This was actually very important in order to realise the detector angles and sizes in 3D, to be provided to the engineers at the mechanical workshops at the Risø and Lund laboratories. Bent Herskind and Gudrun Hagemann later became honorary doctors at the university in Lund in recognition of their important contributions. Later on, the NORBALL concept was further developed into the EUROBALL setup which became a project on a much larger international scale.

## 5 Reflections and reviews

The early experimental and theoretical developments of the work around the back-bending phenomenon are described in [20], and in a contribution to "Advances in Nuclear Physics" [12]. Later developments are described in [15] and in the illuminating review [14]. The back-bending phenomenon is still attracting much interest as shown recently, for example, by Cederwall and collaborators who investigated the exotic $N = Z$ nucleus $^{88}$Ru, and found that the back-bending frequency increases for neutron-deficient nuclei [5].

I particularly want to recommend the summary [21] whose interpretation of the experimental data involves the effect of Coriolis interaction on the nucleons that can be treated within the particle-plus-rotor model (PPR) including the Bardeen–Cooper–Schrieffer (BCS) pairing effects. In this connection, I would like to refer to the very pedagogical mechanical analogue by Mark Riley [16]. Stephens and Lee [21] also have a beautiful presentation of a detailed treatment of the experimental measurements for the odd-mass Er isotopes that we had presented in [8]. In a concluding remark, they pointed out that an important consequence of the discovery was the increased interest in the study of rapid rotation of nuclei as a way to understand nuclear structure.

In hindsight, I think a key point that made this discovery possible was indeed the project's main goal of the systematic investigation of the different elements, an undertaking that was inspired by the similar endeavours earlier performed. The tedious work of systematic mapping turned out to be very rewarding. This point had not been sufficiently considered in the experiments at the Berkeley Laboratory, which I think was the main reason that the phenomenon was overlooked in their experiments. Another key point was the accuracy of our measurements of the ground-state rotational energy levels, facilitated by using the new Ge crystals.

Finally, I have always been fascinated by the fact that similar structural changes in fast rotating matter are also observed in the Universe. Neutron stars are sometimes observed to abruptly increase their rotational velocity due to internal processes, by so-called glitches, probably due to star-quakes. It is really stimulating to ponder that such structural changes occur also on such a very different scale from that of our atomic nuclei.

**Acknowledgements** I thank the referees for useful comments and Felix Ryde for his generous and insightful support.

**Funding** Open access funding provided by Lund University.